\documentclass[12pt]{article}

\usepackage[mathscr]{eucal}
\usepackage[utf8]{inputenc}
\usepackage{slashed}
\usepackage{subcaption}
\usepackage{feynmp}
\usepackage{tikz-feynman}
\tikzfeynmanset{compat=1.1.0}
\usepackage{slashed}
\usepackage{amssymb,amsmath}

\usepackage{slashed}

\newcommand{\bea}{\begin{eqnarray}}
\newcommand{\eea}{\end{eqnarray}}

\newcommand{\be}{\begin{equation}}
\newcommand{\ee}{\end{equation}}
\newcommand{\en}{\end{equation}}
\newcommand{\ba}{\begin{eqnarray}}
\newcommand{\ea}{\end{eqnarray}}

\begin{document}

\begin{titlepage}

\begin{flushright}
arXiv:2211.11399\\
\end{flushright}

\begin{center}
{\Large \bf Three- and Four-Point Functions in CPT-Even\\
Lorentz-Violating Scalar QED}
\end{center}

\begin{center}
{\large B. Altschul,$^{1}$\footnote{{\tt altschul@mailbox.sc.edu}}
L. C. T. Brito,$^{2}$\footnote{{\tt lcbrito@ufla.br}}
J. C. C. Felipe,$^{3}$\footnote{{\tt jean.cfelipe@ufvjm.edu.br}}
S. Karki,$^{1}$\footnote{{\tt karkis@email.sc.edu}}\\
A. C. Lehum,$^{4}$\footnote{{\tt lehum@ufpa.br}}
A. Yu. Petrov,$^{5}$\footnote{{\tt petrov@fisica.ufpb.br}}}

\vspace{5mm}
$^{1}${\sl Department of Physics and Astronomy,} \\
{\sl University of South Carolina, Columbia, SC 29208} \\
$^{2}${\sl Departamento de F\'{i}sica, Instituto de Ci\^{e}ncias Naturais} \\
{\sl Universidade Federal de Lavras, Caixa Postal 3037, 37200-900, Lavras, MG, Brasil} \\
$^{3}${\sl Instituto de Engenharia, Ci\^{e}ncia e Tecnologia,} \\
{\sl Universidade Federal dos Vales do Jequitinhonha e Mucuri, Avenida Um} \\
{\sl 4050, 39447-790, Cidade Universit\'{a}ria, Jana\'{u}ba, MG, Brazil} \\
$^{4}${\sl Faculdade de F\'{i}sica, Universidade Federal do Par\'{a}, 66075-110, Bel\'{e}m, PA, Brazil} \\
$^{5}${\sl Departamento de F\'{i}sica, Universidade Federal da Para\'{i}ba,} \\
{\sl Caixa Postal 5008,	58051-970 Jo\~{a}o Pessoa, PB, Brazil}

\end{center}

\medskip

\centerline {\bf Abstract}

\bigskip

The renormalization of quantum field theories usually assumes Lorentz and gauge symmetries,
besides the general restrictions imposed by unitarity and causality. However, the set of
renormalizable theories can be enlarged by relaxing some of these assumptions. In this work, we
consider the particular case of a CPT-preserving but Lorentz-breaking extension of scalar QED.
For this theory, we calculate the one-loop radiative corrections
to the three- and four-point scalar-vector vertex functions, at the lowest
order in the Lorentz violation parameters; and we explicitly verify that the resulting low-energy
effective action is compatible with the usual gauge invariance requirements. With these results,
we complete the one-loop renormalization of the model at the leading order in the Lorentz-violating
parameters.

\bigskip

\end{titlepage}

\newpage

\section{Introduction}

Studies of Lorentz symmetry breaking represent an important line of research within the quantum
field theory. The modern starting point of this line has been given by the foundational
papers, Refs.~\cite{kostelecky1,kostelecky2}, in which the Lorentz-violating (LV) standard model extension (SME)
was formulated. In Ref.~\cite{kostelecky2}, the complete action of the minimal SME, which is
an effective field theory that includes gauge, spinor, and scalar sectors, was been written down,
and some of the first examples of perturbative calculations were presented. The SME Lagrange density
includes vector- and tensor-valued objects constructed out of existing quantum fields, which are
contracted with background objects that represent favored spacetime direction structures.
However, since the development of the theory,
radiative corrections to the tree-level SME have been considered mostly in the context of spinorial
quantum electrodynamics (QED) and its non-Abelian generalizations. (The results for lower-order
radiative corrections in the minimal LV QED can be found in \cite{KosPic}, for a general review on
radiative corrections in spinorial LV QED see also Ref.~\cite{ourrev}
and references therein.)
 
There have been relatively few papers treating perturbative aspects of Lorentz violation
in scalar field theories,
including LV scalar QED. Among the key papers in the development of the the scalar sector of the SME,
we point especially to Ref.~\cite{higgs}, which examined the modifications to the LV Abelian Higgs model;
Ref.~\cite{HiggsLV}, in which the Higgs mechanism
in LV scalar QED was further studied, with the additional inclusion of a CPT-violating Carroll-Field-Jackiw
(CFJ) term and an analysis of the one-loop corections; Ref.~\cite{Alt1}, which looked at tree-level corrections
to fermion scattering in LV Yukawa theory; Ref.~\cite{Reyes}, where one-loop corrections in a scalar
field theory with additional higher-derivative Myers-Pospelov-like LV terms were
evaluated; Ref.~\cite{Fur}, in which perturbative calculations in a simplified extension of the scalar QED
sector of the SME were performed; and Ref.~\cite{ourEP}, where one-loop effective potentials for various
LV scalar field theories were calculated. 

The study of field theories with broken Lorentz symmetry is not merely a subject of abstract theoretical
interest. The real possibility that there could be tiny deviations from perfect Lorentz invariance
has been examined experimentally, in a wide variety of physical systems and using a remarkable array of
different experimental techniques~\cite{table}. Active areas of experimental interest include
measurements with gravitational waves~\cite{gravwave1,gravwave2,gravwave3}, cosmic rays~\cite{rays2,rays1},
neutrinos, and gamma-ray burst photons~\cite{burst}. In the context of multi-messenger astronomy, for example,
we may have exciting new possibilities for testing and discriminating different models with Lorentz violation
in the near future~\cite{multi}. Because of the interest in these experimental tests, having a complete and
accurate understanding of the relevant sectors of the SME is quite important, and a key part of understanding
the theory is understanding the consequences of quantum corrections. Calculations of 
radiative corrections can play important roles in setting strong, reliable limits on LV
background tensors, or in interpreting evidence of physical Lorentz violation if it is ever uncovered.

In this paper, we will be looking at the scalar QED sector of the SME, including LV modifications not only
in the scalar sector as it has been done in Ref.~\cite{Fur}, but in the gauge sector as well. In both sectors,
the LV terms are CPT-even, having the standard SME forms $c^{\mu\nu}(D_{\mu}\phi)^*D_{\nu}\phi$ and
$\frac{1}{4}\kappa^{\mu\nu\rho\sigma}F_{\mu\nu}F_{\rho\sigma}$~\cite{kostelecky2}. It is interesting to
note that, if the constant tensors $c^{\mu\nu}$ and $\kappa^{\mu\nu\rho\sigma}$ are characterized by only
a single constant background vector $u_{\mu}$, we recover an aether-like theory~\cite{aether}. This
will be a continuation of the previous work~\cite{MPLA2022}, in which the one-loop corrections
to the gauge and scalar two-point functions were explicitly evaluatied. However, here one of our main
aims will be to check the gauge invariance of the quantum corrections in the scalar sector.

The structure of the paper is as follows. In section~\ref{sec-model}, we formulate the CPT-even
LV scalar QED and write down the free propagators. In section~\ref{sec-threepoint},
we calculate the one-loop corrections to the three-point correlation functions, and in
section~\ref{sec-fourpoint} the four-point functions. In section~\ref{sec-eff-action},
the final results are collected, and in section~\ref{sec-RG}, we look at renormalization group
(RG) issues, and
explicitly evaluate the $\beta$-functions for the LV parameters. Finally, our results are summarized
in section~\ref{sec-concl}.

\section{The Model}

\label{sec-model}

Our departure point will be the LV but CPT-even scalar QED Lagrange density (c.f. \cite{MPLA2022}):
\bea
\label{lag1}
{\cal L}=(\eta^{\mu\nu}+c^{\mu\nu})(D_{\mu}\phi)^{*}(D_{\nu}\phi)-m^{2}\phi^{*}\phi
-\frac{1}{4}F^{\mu\nu}F_{\mu\nu}
+\frac{1}{4}\kappa^{\mu\nu\rho\sigma}F_{\mu\nu}F_{\rho\sigma},
\eea
where $D_{\mu}\phi=\partial_{\mu}\phi+ie\phi A_{\mu}$ is the covariant derivative, the metric signature is
$\eta^{\mu\nu} = (1,-1,-1,-1)$, $m^{2} > 0$ is a mass-squared parameter, and $\phi$ the charged scalar field.
Here, in contrast to Ref.~\cite{Fur}, Lorentz symmetry breaking terms appear in both the scalar and gauge 
sectors---via the constant tensors $c^{\mu\nu}$ and $\kappa^{\mu\nu\rho\sigma}$. As usual, we take both
tensors to be  traceless (for $\kappa^{\mu\nu\rho\sigma}$, by this we mean $\kappa^{\mu\nu}{}_{\mu\nu}=0$),
$c^{\mu\nu}$ to be symmetric, and $\kappa^{\mu\nu\rho\sigma}$ to display the
same symmetry as the Riemann curvature tensor. Since $c^{\mu\nu}$ and $\kappa^{\mu\nu\rho\sigma}$ will
generally mix under radiative corrections, we expect that if one of them is present, they generally both must
appear if the theory is to be strictly renormalizable.
Note that we would also need a term $(\phi^{*}\phi)^{2}$ in the Lagrange density~(\ref{lag1}) to guarantee
the full renormalizability of the model. However, we are omitting this term here because quantum corrections to
it will not be  affected by the Lorentz violation; this is a consequence of the tracelessness of the
LV tensors.

The propagators of the theory, in momentum space, have the forms
\bea
G(k) &=& \langle\phi(k)\phi^*(-k)\rangle = \frac{i}{k^{2}-m^{2}+i\epsilon}
-\frac{ic^{\mu\nu}k_{\mu}k_{\nu}}{(k^{2}-m^{2}+i\epsilon)^{2}}
\label{scalar_propag} \\ 
\Delta^{\mu\nu}(k) &=& \langle A^{\mu}(k)A^{\nu}(-k)\rangle=
-\frac{i\eta^{\mu\nu}}{k^{2}+i\epsilon}
+\frac{2i\kappa^{\mu\rho\nu\sigma}k_{\rho}k_{\sigma}}{(k^{2}+i\epsilon)^{2}}
\label{props}.
\eea
For convenience, we have employed the Feynman gauge---adding the usual Lorentz-invariant gauge-fixing term
$-\frac{1}{2}(\partial^{\mu}A_{\mu})^{2}$ to~(\ref{lag1}).  We have then expanded the gauge propagator up to the
first order in the LV parameters $c^{\mu\nu}$ and $\kappa^{\mu\rho\nu\sigma}$. As in Ref.~\cite{MPLA2022} and
as is commonplace in the literature, we are only taking into account the leading-order contributions from
the LV background tensors. This is justified by the physical observation that Lorentz violation, if it exists
is known to be a small effect, so it makes sense to treat it as a perturbation on top of the well-understood
Lorentz-invariant theory.

Both of the CPT-even terms with $c^{\mu\nu}$ and $\kappa^{\mu\rho\nu\sigma}$ were 
introduced originally in Ref.~\cite{kostelecky2}. Because the coefficients $c^{\mu\nu}$ and
$\kappa^{\mu\rho\nu\sigma}$ are dimensionless, it is expected based on power counting that the modified, LV
scalar QED theory will continue to be renormalilzable. The tensor $c^{\mu\nu}$ of the present paper should
not be confused with the one normally introduced in the fermionic sector of SME (although the two
affect the dispersion relations for the scalar and fermion fields in homologous ways).
What we are calling $c^{\mu\nu}$ was the $k^{\mu\nu}_{\phi\phi}$ tensor of~\cite{kostelecky2},
which can, most generally, possess a symmetric real part and an antisymmetric imaginary part.
However, since we are considering here only a real $c^{\mu\nu}$, it must necessarily be symmetric.

Along with the modified propagators coming from the bilinear part of the Lagrange density~(\ref{lag1}),
there are interactions, which are also modified by the presence of the $c^{\mu\nu}$ term. The vertices arise
out of the presence of the covariant derivative $D_{\mu}\phi$ (which involves $A_{\mu}$) in the action.
It is a consequence of gauge
invariance that same quantity $c^{\mu\nu}$ must appear in both the free scalar propagator and
the three- and four-field gauge-scalar vertices.
In the next two sections we will be evaluating quantum correction involving the tree-level vertices 
\begin{equation}
V_{3}=i(\eta^{\mu\nu}+c^{\mu\nu})A_{\mu}(\phi\partial_{\nu}\phi^{*}-\phi^{*}\partial_{\nu}\phi)
\label{V3}
\end{equation}
and 
\begin{equation}
V_{4}=(\eta^{\mu\nu}+c^{\mu\nu})A_{\mu}A_{\nu}\phi\phi^{*}.
\label{V4}
\end{equation}
(We shall generally drop the coupling constant $e$, although it will be restored at the end of
our analysis.)
In order to evaluate all the relevant Green's functions, we shall use an adapted version of a set of
Mathematica packages~\cite{feyncalc,feynarts,feynrules}.

\section{Results for the Three-Point Function}

\label{sec-threepoint}

\begin{figure}[h]
	\begin{center}
		\includegraphics[scale=0.9]{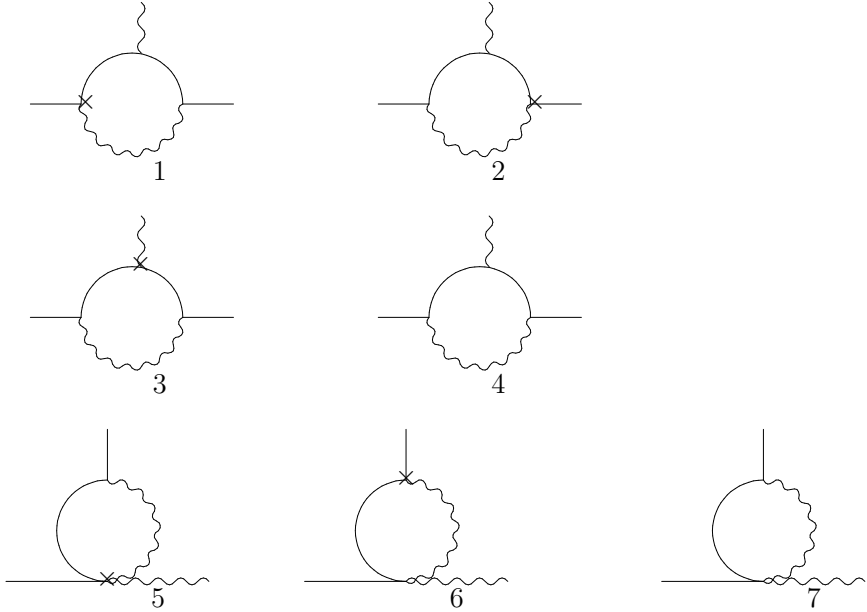}
		\caption{Contributions to three-point gauge-scalar vertex. The wavy and solid lines represent
		the photon and scalar propagators, respectively, and the crosses indicate insertions of
		LV $c^{\mu\nu}$ parameters at the vertices. The usual Lorentz-invariant contributions come from
		portions of diagrams 4 and 7.}
		\label{Fig1}
	\end{center}
\end{figure}

Now, we turn to the Feynman diagrams contributing to the three-point vertex function 
$\langle A\phi\phi^{*}\rangle$,  which gives the quantum corrections to the vertex (\ref{V3}).
All internal propagators now are ``dressed,'' so that they depending on the LV parameters,
given by (\ref{scalar_propag}) and (\ref{props}).  We obtain the Feynman rules for the vertices
as usual from (\ref{V3}) and (\ref{V4}). The corresponding graphs are depicted in figure~\ref{Fig1}.   

The contributions from the diagrams numbered 1--7 have the respective forms given by eqs.
(\ref{eq-I1}--\ref{eq-I7})
(with $p_{3}=-p_{1}-p_{2}$ from the conservation of the total incoming external momentum),
\begin{eqnarray}
I_{1} = & i \phi(p_{1})\phi^{*}(p_{2})A^{\mu}(p_{3}) c^{\nu\rho} & \int \frac{d^{4}k}{(2\pi^{4})}\,
G(k)G(k+p_{3})\Delta_{\nu\sigma}(k-p_{1}) \nonumber\\
& & \quad \times (2k+p_{3})_{\mu}(k+p_{1})_{\rho}(k+p_{3}-p_{2})^{\sigma} 
\label{eq-I1} \\
I_{2} = & i\phi(p_{1})\phi^{*}(p_{2})A^{\mu}(p_{3}) c^{\nu\rho} & \int \frac{d^{4}k}{(2\pi^{4})}\,
G(k)G(k+p_{3})\Delta_{\sigma\nu}(k-p_{1}) \nonumber\\
& & \quad \times (2k+p_{3})_{\mu}(k+p_{1})^{\sigma}(k+p_{3}-p_{2})_{\rho} \\
I_{3}= & i \phi(p_{1})\phi^{*}(p_{2})A^{\mu}(p_{3}) c_{\mu}{}^{\nu} & \int \frac{d^{4}k}{(2\pi^{4})}\,
G(k)G(k+p_{3})\Delta^{\rho\sigma}(k-p_{1}) \nonumber\\
& & \quad \times (2k+p_{3})_{\nu}(k+p_{1})_{\rho}(k+p_{3}-p_{2})_{\sigma} \\
I_{4}= & i \phi(p_{1})\phi^{*}(p_{2})A^{\mu}(p_{3}) & \int\frac{d^{4}k}{(2\pi^{4})}\,
G(k)G(k+p_{3})\Delta^{\rho\sigma}(k-p_{1}) \nonumber\\
& & \quad \times (2k+p_{3})_{\mu}(k+p_{1})_{\rho}(k+p_{3}-p_{2})_{\sigma},
\end{eqnarray}
for the four diagrams with only three-field vertices internally and, for the last three diagrams
(each of which includes a four-field vertex),
\begin{eqnarray}
I_{5} & = & 2i\left[\phi(p_{1})\phi^{*}(p_{2})-\phi^{*}(p_{1})\phi(p_{2})\right]A^{\mu}(p_{3})c_{\mu}{}^{\nu}
\int\frac{d^{4}k}{(2\pi^{4})}\, \Delta_{\nu\rho}(k+p_{2})(k-p_{2})^{\rho}G(k)\quad \\
I_{6} & = & 2i\left[\phi(p_{1})\phi^{*}(p_{2})-\phi^{*}(p_{1})\phi(p_{2})\right]A^{\mu}(p_{3})c^{\nu\rho}
\int\frac{d^{4}k}{(2\pi^{4})}\, \Delta_{\mu\nu}(k+p_{2})(k-p_{2})_{\rho}G(k) \\
I_{7}&=&2i\left[\phi(p_{1})\phi^{*}(p_{2})-\phi^{*}(p_{1})\phi(p_{2})\right]A^{\mu}(p_{3})
\int\frac{d^{4}k}{(2\pi^{4})}\, \Delta_{\mu\nu}(k+p_{2})(k-p_{2})^{\nu}G(k).
\label{eq-I7}
\end{eqnarray}
The only diagrams that do not involve LV tensors at the vertices are $I_{4}$ and $I_{7}$. So the
Lorentz-violating contributions for these diagrams arise solely from the Lorentz-violating dressing
of the propagators (\ref{scalar_propag}) and (\ref{props}). In contrast, in the diagrams $I_{1}$--$I_{3}$
and $I_{5}$--$I_{6}$, we need only keep Lorentz-invariant parts of the propagators, since the
vertices ensure that the diagrams already depend on $c^{\mu\nu}$.

In order to calculate the vertex contributions $I_{1}$ through $I_{7}$, we may use the Implicit Regularization
(IR) framework to isolate the divergent parts of amplitudes. (See Refs.~\cite{ref-IR1,ref-IR2,ref-IR3}
for general reviews of the
method.) Here we are interested only in the divergences of the radiative corrections, and thus we shall
express the results  in terms of the single logarithmically divergent integral
\begin{equation}
\label{ilog}
I_{\log}(m^2)=\int^{\Lambda} \frac{d^{4}k}{(2\pi)^{4}}\frac{1}{(k^{2}-m^{2})^{2}}.
\end{equation}
The $\Lambda$ annotation indicates that the integral is regularized in some manner compatible with the
gauge symmetry. More precisely, this assumption is equivalent to adopting $\alpha_{i}=0$ ($i=1,2$) in the
general relations
\begin{eqnarray}
\label{supef1}
\int^\Lambda \frac{d^{4}k}{(2\pi)^{4}}\frac{k_{\mu} k_{\nu}}{(k^{2}-m^{2})^{3}} & = & 
\frac{\eta_{\mu\nu}}{4}\left[I_{\log}(m^2) + \alpha_{1}\right] \\
\label{supef2}
\int^\Lambda \frac{d^{4}k}{(2\pi)^{4}}\frac{k_{\mu}k_{\nu} k_{\rho}k_{\sigma}}{(k^{2}-m^{2})^{4}} & = &
\frac{\eta_{\mu\nu}\eta_{\rho\sigma}+\eta_{\mu\rho}\eta_{\nu\sigma}+\eta_{\mu\sigma}\eta_{\nu\rho}}{24}
\left[I_{\log}(m^2) + \alpha_2\right].
\end{eqnarray}
The IR method may also be applied to the naive quadratic divergences in scalar QED
theories, replacing those formally divergent integrals with alternative expressions,
in such a way that complete cancellation of all potential quadratic divergences is assured
and transversality of the gauge theory is maintained. This entails
setting the analogous $\alpha$ coefficients in the corresponding formal expressions
for the integrals with power-counting quadratic divergences to be equal to zero. 

Evaluating the integrals (\ref{eq-I1}--\ref{eq-I7}) is fairly tedious. After a lengthy calculation,
we find that
\begin{eqnarray}
I_{1} = I_{2} & = & \left[-\frac{2i}{3}c^{\mu\nu}(p_{1}-p_{2})_{\nu}\right]I_{\log}(m^{2})
\phi(p_{1})\phi^{*}(p_{2})A_{\mu}(p_{3})
\label{i1} \\
I_{3} & = & \left[-ic^{\mu\nu}(p_{1}-p_{2})_{\nu}\right]
I_{\log}(m^{2})\phi(p_{1})\phi^{*}(p_{2})A_{\mu}(p_{3}) \\
I_{4} & = & \left[\frac{2i}{3}c^{\mu\nu}(p_{1} -p_{2})_{\nu}\right]I_{\log}(m^{2})\phi(p_{1})\phi^{*}(p_{2})A_{\mu}(p_{3}) \\
I_{5}=I_{6} & = & \left[3ic^{\mu\nu}(p_{1}-p_{2})_{\nu}\right]I_{\log}(m^{2})\phi(p_{1})\phi^{*}(p_{2})A_{\mu}(p_{3}) \\
I_{7} & = & \left[-i\left(\frac{1}{3}c^{\mu\nu}+\kappa^{\mu\rho\nu}{}_{\rho}
\right)(p_{1}-p_{2})_{\nu}\right]I_{\log}(m^{2})\phi(p_{1})\phi^{*}(p_{2})A_{\mu}(p_{3})
\label{i7}.
\end{eqnarray}
Each of these expressions has the structure of the momentum transfer at the vertex $p_{1}-p_{2}$,
contracted with a two-index symmetric tensor constructed out of the Lorentz-violating backgrounds. The
terms in which the tensors that appear are simply multiples of $c^{\mu\nu}$ will contribute to the
self-renormalization of the $c^{\mu\nu}$ tensor. However, the presence of $\kappa^{\mu\rho\nu}{}_{\rho}$
in (\ref{i7}) demonstrates explicitly that there is renormalization mixing between the background tensors
from the gauge and scalar sector.
On the other hand, it is interesting to note that the formula for for $I_{4}$ does not depends on
$\kappa^{\mu\nu\rho\sigma}$, since this background tensor is antisymmetric under
the exchanges of the first two and of the last two indices.

In parallel with the Mathematica calculations, we also
verified these expressions by hand. The manual calculations set the explicit integral
expression for each diagram and used dimensional regularization (DR) to evaluate them.

These results (\ref{i1}--\ref{i7}) will be employed as part of the determination of the full
gauge-scalar contribution to the one-loop effective action in section~\ref{sec-eff-action}. Note that
the RG $\beta$-function for the charge Lorentz-violating vertex
coefficient may be determined solely from the
leading quantum correction to the three-field vertex amplitude,
in conjunction with the one-loop gauge and scalar self-energies. However,
in order to confirm that the renormalization-improved theory remains gauge invariant, we also need
to calculate the lowest-order radiative corrections to the four-field vertex.

\section{Results for the Four-Point Function}

\label{sec-fourpoint}

So the next step in our analysis will be calculating the radiaive contributions to the 
four-point function $\langle AA\phi\phi^{*}\rangle$, which corrects the vertex (\ref{V4}). Note that
the tree-level version of the vertex does not contain any derivatives.

\begin{figure}[h]
	\begin{center}
		\includegraphics[scale=0.9]{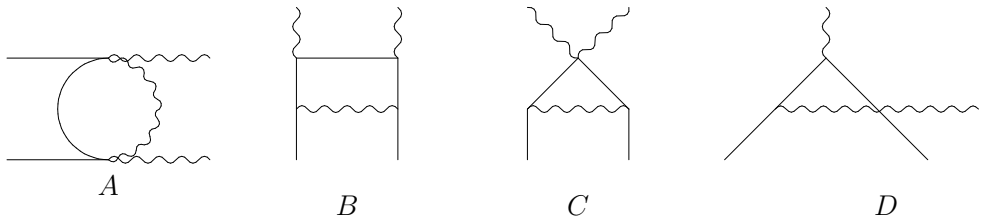}
		\caption{Contributions to four-point gauge-scalar functions.}
		\label{Fig2}
	\end{center}
\end{figure}

In this case  the Feynman diagrams are those shown in figure~\ref{Fig2}. The analytical expressions
for the potentially divergent parts of diagrams A--D are
\begin{eqnarray}
I_{A} & = & -2\phi(p_{1})\phi^{*}(p_{2})A^{\mu}(p_{3})A^{\nu}(p_{4})(\eta_{\mu\rho}+c_{\mu\rho})
(\eta_{\nu\sigma}+c_{\nu\sigma})\int\frac{d^{4}k}{(2\pi)^{4}}\,\Delta^{\rho\sigma}(k)G(k) \\
I_{B} & = & 2\phi(p_{1})\phi^{*}(p_{2})A^{\mu}(p_{3})A^{\nu}(p_{4})(\eta_{\mu\rho}+c_{\mu\rho})
(\eta_{\nu\sigma}+c_{\nu\sigma})(\eta_{\kappa\lambda}+c_{\kappa\lambda})(\eta_{\tau\varphi}+c_{\tau\varphi})
\nonumber \\
& & \times\int\frac{d^{4}k}{(2\pi)^{4}}\left[G(k)\right]^{3}\Delta^{\kappa\tau}(k)
k^{\rho}k^{\sigma}k^{\lambda}k^{\varphi} \\
I_{C} & = & -2\phi(p_{1})\phi^{*}(p_{2})A^{\mu}(p_{3})A^{\nu}(p_{4})(\eta_{\mu\nu}+c_{\mu\nu})
(\eta_{\rho\sigma}+c_{\rho\sigma})(\eta_{\kappa\lambda}+c_{\kappa\lambda})
\nonumber\\
& & \times \int\frac{d^{4}k}{(2\pi)^{4}}\left[G(k)\right]^{2}\Delta^{\rho\kappa}(k)k^{\sigma}k^{\lambda} \\
I_{D} & = & 2\phi(p_{1})\phi^{*}(p_{2})A^{\mu}(p_{3})A^{\nu}(p_{4})(\eta_{\mu\rho}+c_{\mu\rho})
(\eta_{\nu\sigma}+c_{\nu\sigma})(\eta_{\kappa\lambda}+c_{\kappa\lambda}) \nonumber\\
& & \times \int\frac{d^{4}k}{(2\pi)^{4}}\left[G(k)\right]^{2}\Delta^{\lambda\sigma}(k)k^{\rho}k^{\kappa}.
\end{eqnarray}
Note that in obtaining these expressions, we have neglected all external momenta. They may be freely
set to zero in the determination of the infinite renormalization of (\ref{V4}); this is related
to the fact that the vertex $V_{4}$ is momentum independent. The naive degree of divergence of the diagrams
shown in figure~\ref{Fig2} is zero, so any appearance of a factor of an external leg momentum in
the numerator of one of the integrals involved would render the integral involved finite---and thus
negligile, since we are only considering the effects of the formally divergent radiative corrections.

These contributions must be expanded up to first orders in SME parameters. That is, we must
again use the propagators expanded up to the first order in $c^{\mu\nu}$ and $\kappa^{\mu\nu\rho\sigma}$.
Proceeding in the same fashion as for the three-point functions, we have found the following results for
the four-point vertex corrections:
\begin{eqnarray}
I_{A} & = & \left(4c^{\mu\nu}+\kappa^{\mu\rho\nu}{}_{\rho}\right)I_{\log}(m^{2})
\label{A} \\
I_{B} & = & -\frac{5}{6}c^{\mu\nu}I_{\log}(m^{2}) \\
I_{C} & = & 2c^{\mu\nu}I_{\log}(m^{2}) \\
I_{D} & = & -\frac{7}{6}c^{\mu\nu}I_{\log}(m^{2}). \label{D}
\end{eqnarray}
(Note that the contributions from diagrams B--D actually sum to zero, leaving $I_{A}$ as the sole
contributor to the renormalization of the four-field vertex. However,
while this observation seems potentially suggestive, it is actually specific to
the Feynman gauge and does not hold more generally.)
With these expressions, we have all we need to verify the gauge invariance of the model at the one-loop
level.

\section{Final Results for the Low-Energy Effective Action}

\label{sec-eff-action}

In this section, we shall put together the results for the theory's two-point ($\langle\phi\phi^{*}\rangle$),
three-point ($\langle A\phi\phi^{*}\rangle$), and four-point ($\langle AA\phi\phi^{*}\rangle$)
correlation functions. This will enable us to perform an explict verification of the gauge invariance
of the renormalized effective action, at first order in the loop expansion and likewise first order in
the Lorentz violating parameters $c^{\mu\nu}$ and $\kappa^{\mu\nu\rho\sigma}$.
The expression for the two-point function was previously derived in Ref.~\cite{MPLA2022}, whereas
the results for the three- and four-point functions have been calculated in the two preceding sections;
the ultimate contributions from the diagrams in figures~\ref{Fig1} and~\ref{Fig2} are given by
(\ref{i1}--\ref{i7}) and (\ref{A}--\ref{D}).

In order to maintain gauge invariance (and thus renormalizability and unitarity) at the pertubative order
that we are interested in, the infinte parts of $\langle\phi\phi^{*}\rangle$, $\langle A\phi\phi^{*}\rangle$,
and $\langle AA\phi\phi^{*}\rangle$ need to
satisfy certain relations. The radiatively-generated contributions
to the effective action must, when taken together, have the same structure as the covariant derivative term
in the original action. The net contributions must assemble to form an expression of the form
$k^{\mu\nu}(D_{\mu}\phi)^{*}(D_{\nu}\phi)$, with some (logarithmically divergent) constant tensor $k^{\mu\nu}$.
In momentum space, this kind of term takes the form
\begin{equation}
\label{covariant}
k^{\mu\nu}\left[-p_{1\mu}p_{2\nu}-i(p_{1}-p_{2})_{\mu}A_{\nu}(p_{3})
+A_{\mu}(p_{3})A_{\nu}(p_{4})\right]\phi(p_{1})\phi^{*}(p_{2}).
\end{equation}
In fact, the sum of all diagrams in figures~\ref{Fig1} and~\ref{Fig2}, as well as the
scalar self-energy diagrams~\cite{MPLA2022} takes precisely this form. The divergent part $\Gamma_{{\rm div}}$
of the resulting sum looks like
\begin{equation}
\label{result}
\Gamma_{{\rm div}} = \left(4c^{\mu\nu}+\kappa^{\mu\rho\nu}{}_{\rho}\right)I_{\log}(m^{2})
\left[-p_{1\mu}p_{2\nu}-i(p_{1}-p_{2})_{\nu}A_{\mu}(p_{3})+A_{\mu}(p_{3})A_{\nu}(p_{4})\right]
\phi(p_{1})\phi^{*}(p_{2}),
\end{equation}
which clearly matches (\ref{covariant}).

\section{Renormalization Group Functions}

\label{sec-RG}

So far, we have derived the divergent part of the effective action for the model (\ref{lag1}).
In this section, we shall compute the RG functions associated with the theory, in particular
the $\beta$-functions that describe the dependences of the SME terms on the interaction scale.
For clarity, we shall take the $c^{\mu\nu}$ and $\kappa^{\mu\nu\rho\sigma}$ tensors to have
particularly simple forms, so that the RG scaling for each of them may be reduced to the behavior of
a single scalar quantity. However, it would be a completely straightforward generalization to separate out
the individual $\beta$-functions for the individual Lorentz components of the tensors.

The specific form we shall assume for the matter-sector SME coefficients is
\begin{equation}
c^{\mu\nu}=Q_{c}u^{\mu}u^{\nu},
\label{eq-Q-c}
\end{equation}
where $u^{\mu}$ is a fixed unit(-like) dimensionless four-vector. A theory in which
all the LV backgrounds depend on just a single such four-vector are (especially when the vector in
question is purely timelike) often referred to as ``aether-like'' LV theories. In the context of spontaneous
breaking of Lorentz symmetry, models with just a single preferred four-vector background are 
``bumblebee'' models~\cite{ref-kost12}. The bumblebee framework involves a single dynamical four-vector
field acquiring a vacuum expectation value, which sets the spacetime direction of $u^{\mu}$. In principle,
the field could be timelike, spacelike, or lightlike (depending, for example, on the structure of the
potential responsible for spontaneous Lorentz symmetry breaking). However, the expression~(\ref{eq-Q-c})
is subtly defective if $u^{2}\neq 0$, since in that case the trace $c^{\mu}{}_{\mu}$ is nonzero, contrary
to standard conventions. It would be quite straightforward to rectify this problem by subtracting an
additional diagonal tensor from (\ref{eq-Q-c}). However, in the interest of maintaining maximal simplicity
in our calculations, we shall instead assume that $u^{\mu}$ is simply lightlike---$u^{2}=0$ implying that
$c^{\mu\nu}$ is traceless.

The Lorentz violation coefficient $\kappa^{\mu\nu\rho\sigma}$ in the pure electromagnetic sector will
also be taken to depend on just the lightlike $u^{\mu}$ and an overall normalization constant. The
specific form is
\begin{equation}
\label{eq-Q-kappa}
\kappa^{\mu\nu\rho\sigma}=Q_{\kappa}(u^{\mu}u^{\rho}\eta^{\nu\sigma}-u^{\mu}u^{\sigma}\eta^{\nu\rho}
+u^{\nu}u^{\sigma}\eta^{\mu\rho}-u^{\nu}u^{\rho}\eta^{\mu\sigma}).
\end{equation}
Like $Q_{c}$, $Q_{\kappa}$ is dimensionless.
When we need to denote the bare versions of  quantities, we shall include an additional
subscript $0$, so that $Q_{\kappa 0}$ and $Q_{c0}$ stand for the bare parameters which appear in 
bare version of the Lagrange density (\ref{lag1}).

Although it is possible to calculate the RG functions using the IR
formalism directly~\cite{Brito:2008zn}, in this section we will express our results in the more
commonly used language of DR. It is actually simple to carry expressions over from IR to DR. In IR,
the RG scale $\mu$ is introduced via the identity
\begin{equation}
\label{regsacale}
I_{\log}(m^{2}) - I_{\log}(\mu^{2}) = \frac{i}{16\pi^2} \ln\left(\frac{\mu^{2}}{m^{2}}\right). 
\end{equation}
To obtain the same results as in
conventional DR, we may just set all the $\alpha_{i}$ parameters to zero in (\ref{supef1}) and (\ref{supef2}),
use the identity (\ref{regsacale}) to write the Green's functions as a function of $\mu$,
and substitute for the $I_{\log}(\mu^{2})$ defined in (\ref{ilog}) the DR formula 
\begin{equation}\label{formula}
\mu^{D-4}\int\frac{d^{D}k}{(2\pi)^{D}}\frac{1}{(k^{2}-\mu^{2})^{2}}=
\frac{i}{16\pi^{2}}\left(\frac{1}{\epsilon}+\ln 4\pi-\gamma\right)+\mathcal{O}(\epsilon).
\end{equation}
Here, we defined as usual $\epsilon = \frac{D-4}{2}$, where $D$ is the analytically
continued dimension of spacetime. In what follows, we will be interested only in the singular term containing
$1/\epsilon$ in (\ref{formula}).

We define the renormalized fields $\phi_{0}=Z_{2}^{1/2}\phi$ and $A^{\mu}_{0}= Z_{3}^{1/2}A^{\mu}$ and
rewrite the Lagrangian  (\ref{lag1})---taken to depend on the bare fields---in terms of the renormalized
fields (and at this stage, we also restore the previously omitted coupling constant $e$, which
is equivalent to taking the bare charge to be $e_{0}=1$)
\begin{eqnarray}
\label{eq04}
\mathcal{L} & = &  Z_{2}(\partial^{\mu}\phi^{*})(\partial_{\mu}\phi) - m^{2}Z_{m}\phi^{*} \phi
-\frac{1}{4}Z_{3}F^{\mu\nu}F_{\mu\nu}+ie Z_{1} A^{\mu}\left(\phi^{*} \partial_{\mu} \phi
-\phi\partial_{\mu} \phi^{*}\right) \nonumber\\
& & +{}\, e^{2}Z_{4} A^{\mu}A_{\mu}\phi^{*}\phi+u^{\mu}u^{\nu}\left[Q_{c}Z_{5}
(\partial_{\mu}\phi^{*})(\partial_{\nu}\phi)+ Q_{\kappa} Z_{6}F^{\mu\alpha}F^{\nu}{}_{\alpha}\right] \nonumber\\
& & +{}\,u^{\mu}u^{\nu}\left[ieQ_{c}Z_{7}A_{\nu}\left(\phi^{*}\partial_{\mu}\phi
-\phi\partial_{\mu}\phi^{*}\right)
+ e^{2}Q_{c}Z_{8}\phi^{*}\phi A_{\mu}A_{\nu} \right] +\mathcal{L}_{GF}.
\end{eqnarray}
Here, $\mathcal{L}_{GF}$ is the gauge-fixing term, and the relations among the renormalization constants
are
\begin{eqnarray}
\label{eq-Zm}
m^{2}Z_{m} & = & \mu^{-2\epsilon}m_{0}^{2}Z_{2} \\
eZ_{1} & = & \mu^{-2\epsilon}Z_{2}Z_{3}^{1/2} \\
e^{2}Z_{4} & = & \mu^{-2\epsilon}Z_{2}Z_{3} \\
Q_{c}Z_{5} & = & \mu^{-2\epsilon}Q_{c0}Z_{2} \\
Q_{\kappa}Z_{6} & = & \mu^{-2\epsilon}Q_{\kappa0}Z_{3} \\
eQ_{c}Z_{7} & = & \mu^{-2\epsilon}Q_{c0}Z_{2}Z_{3}^{1/2} \\
e^{2}Q_{c}Z_{8} & = & \mu^{-2\epsilon}Q_{c0}Z_{2}Z_{3}.
\label{eq-Z8}
\end{eqnarray}
The definitions of the renormalization constants account for the fact that the bare Lagrangian
effectively had $e_{0}=1$.
Each of the renormalization constants $Z_{i}$ may be expanded as a power series in the coupling constants
and determined sequentially, order by order in the perturbative expansion---that is,
\begin{eqnarray}
Z_i=1+Z^{(1)}_i+Z^{(2)}_i+\cdots.
\end{eqnarray}

We now need to evaluate the counterterms corresponding to the ultraviolet-divergent parts of the one-loop
corrections to the scalar and photon self-energies (for Feynman diagrams calculated in Ref.~\cite{MPLA2022})
and the vertex functions (given by figures~\ref{Fig1} and~\ref{Fig2} of this paper) in the scalar QED model. 
We start with the one-loop scalar self-energy. The insertion of the LV parameters at the first order
was considered in Ref.~\cite{MPLA2022}. The resulting expression is
\begin{equation}
\label{eq-sse01}
\Sigma_1(p) = \frac{\lambda m^{2}-e^{2} \left(m^{2}+{2} p^{2}\right)}{16\pi^{2}\epsilon}
-\frac{e^{2}(4 Q_{c}+Q_{\kappa})(u\cdot p)^{2}}{16\pi^{2}\epsilon}
+Z_{2}^{(1)}p^{2}-m^{2}Z_{m}^{(1)}+Q_{c}Z_{5}^{(1)}(u\cdot p)^{2},
\end{equation}
from which it is possible to read off what the leading contributions to the renormalization constants
must be,
\begin{eqnarray}
Z_{2}^{(1)} & = & \frac{e^2 }{8\pi^{2}\epsilon}
\label{ct01} \\
Z_{m}^{(1)} & = & \frac{\left(\lambda-e^{2}\right)}{16\pi^{2}\epsilon}
\label{ct02}\\
Z_{5}^{(1)} & = & \frac{e^{2} (4Q_{c}+Q_{\kappa})}{16\pi^{2}Q_{c}\epsilon}.
\label{ct03}
\end{eqnarray}
Note that in (\ref{eq-sse01}--\ref{ct01}), we have tacitly added back in the usual contribution from the
four-scalar coupling $\lambda$. Due to the tracelessness of $c^{\mu\nu}$ and $\kappa^{\mu\nu\rho\sigma}$,
the renormalizations of the usual field strength ($Z_{2}$) and mass ($Z_m$) terms are not affected by
the Lorentz violation. 

The one-loop photon self-energy  with  the LV parameters insertions is similarly given by~\cite{MPLA2022}
\begin{eqnarray}
-i\Pi^{\mu\nu}(p) & = & -\left(p^{2}\eta^{\mu\nu}-p^{\mu}p^{\nu}\right)
\left[\frac{e^{2}}{48\pi^{2}\epsilon}+Z_{3}^{(1)}\right]
+\left\{(u\cdot p)\left[\eta^{\mu\nu}(p\cdot u)-u^{\nu}p^{\mu}\right]\right. \nonumber\\
& & {}\,+ \left.u^{\mu}\left[u^{\nu}p^{2}-(u\cdot p)p^{\nu}\right]\right\}
\left[-\frac{{e}^{2}Q_{c}}{48\pi^2\epsilon}+\frac{Q_{\kappa}Z_{6}^{(1)}}{2}\right].
\end{eqnarray}
The renormalization constants needed to subtract the divergences are
\begin{eqnarray}
\label{1-loop_Z3_sqed}
Z_{3}^{(1)} & = & -\frac{e^{2}}{48\pi^{2} \epsilon} \\
Z_{6}^{(1)} & = & \frac{e^{2} Q_{c}}{24\pi^{2}Q_{\kappa}\epsilon},
\end{eqnarray}
$Z_{3}^{(1)}$ naturally being the usual Maxwell field strength renormalization constant, again
because of the tracelessness of the SME background tensors.

To the older results for the two-point function counterterms, we now add the vertex correction terms,
beginning with the ultraviolet-divergent part of the three-point vertex function.
Using the definition (\ref{eq-Q-kappa}) in the results of section~\ref{sec-threepoint},
and adding the contributions (\ref{i1}--\ref{i7}), we obtain
\begin{equation}
-i\Gamma^{\mu}=e \left(p_{2}^{\mu}-p_{1}^{\mu }\right)\left[\frac{e^{2}}{8\pi^{2}\epsilon}-Z_1^{(1)}\right]
+ e Q_{c}u^{\mu}\left[u\cdot\left(p_{2}-p_{1}\right)\right]
\left[\frac{e^{2}(4Q_{c}+Q_{\kappa})}{16\pi^{2}Q_{c}\epsilon}-Z_{7}^{(1)}\right].
\end{equation}
Imposing finiteness, we have
\begin{eqnarray}
Z_{1}^{(1)} & = & \frac{e^{2}}{8\pi^{2}\epsilon}
\label{eqz1} \\
Z_{7}^{(1)} & = & \frac{e^{2}(4Q_{c}+Q_{\kappa})}{16\pi^{2}Q_{c}\epsilon}.
\label{eqz7}
\end{eqnarray}

The expressions derived so far contain enough information to determine the $\beta$-functions
for $Q_{c}$ and $Q_{\kappa}$. However, to verify the gauge invariance of the renormalized theory, we
also need to look at the four-field vertex corrections (although some of the necessary relations may
already be checked at this stage). Specifically, we need to look at the radiative corrections to the
gauge-scalar four-point function. [As noted above before, we are not interested in the quantum corrections
to the four-field $(\phi^{*}\phi)^2$ vertex, because these corrections
 will not be affected by the Lorentz violation if
$u^{2}=0$.]  Using (\ref{eq-Q-c}--\ref{eq-Q-kappa}) and adding the results (\ref{A}--\ref{D}), we find
\begin{equation}
\Gamma^{\mu\nu}=-\frac{e^{4}\eta^{\mu\nu}}{4\pi^{2}\epsilon}
-\frac{2e^{4}(4Q_{c}+Q_{\kappa})u^{\mu}u^{\nu}}{16\pi^{2}\epsilon}
+2e^{2} Z_{4}^{(1)}\eta^{\mu\nu}+2 e^{2} Q_{c} Z_{8}^{(1)} u^{\mu}u^{\nu},
\end{equation}
which is rendered finite if
\begin{eqnarray}
Z_{4}^{(1)} & = & \frac{e^2}{8\pi^{2}\epsilon}
\label{ctz4} \\
Z_8^{(1)} &=& \frac{e^{2} (4Q_{c}+Q_{\kappa})}{16\pi^{2} Q_{c} \epsilon}.
\label{ctz8}
\end{eqnarray}

Since there are supposed to be, according to (\ref{eq-Zm}--\ref{eq-Z8}) only two underlying divergent
renormalization factors in the theory, we can test the consistency of our calculations with gauge
invariance---for
example, by comparing (\ref{ct01}) with (\ref{eqz1}), and (\ref{ct03}) with (\ref{eqz7}) and
(\ref{ctz8}). The consistency conditions are indeed satisfied, and, in particular, we have
\begin{eqnarray}
Z_{1}^{(1)} & = & Z_{2}^{(1)} \\
Z_{5}^{(1)} & = & Z_{7}^{(1)}=Z_{8}^{(1)},
\label{gauge_four}
\end{eqnarray}
which are the Ward identities for the theory.
In particular, the identities (\ref{gauge_four}) are directly related to the structure we found in
(\ref{result}) for the effective action, and thus they are an explicit manifestations of the
gauge-invariant structure of our result.        

Now we are prepared to compute the RG functions of the model. Based on the relations defining the $Z_{i}$
factors and their specific values obtained in this section, we find the following expressions for
the $\beta$-functions
\begin{eqnarray}
\label{beta_e_sqed1l}
\beta(e) & = & \mu\frac{de}{d\mu}=\frac{e^{3}}{48\pi^{2}} \\
\beta(\lambda) & = & \mu\frac{d\lambda}{d\mu}=\frac{5 \lambda^{2}-12\lambda e^{2}+24e^{4}}{16\pi^{2}}
\label{betae} \\
\beta(Q_{c}) & = & 3\beta(Q_{\kappa})=\frac{e^{2}(2Q_{c}+Q_{\kappa})}{8\pi^{2}}.
\end{eqnarray}
As expected, $\beta(e)$  and $\beta(\lambda)$ are the same as in standard scalar QED without LV terms.  

It may also be interesting to notice some of the consequences of relaxing the assumption
$u^{2}=0$ in the calculations of the RG functions. The anomalous dimensions 
$\gamma_{i}=\frac{1}{2}\frac{d\ln Z_i}{d\ln\mu}$ are given by
\begin{eqnarray}
\gamma_{2} & = & \frac{1}{2}\frac{d\ln Z_{2}}{d\ln\mu}=-\frac{e^{2}}{8\pi^{2}}
+\frac{3e^{2}u^{2} \left(Q_{c}-Q_{\kappa}\right)}{64\pi^{2}} \\
\gamma_{3} & = & \frac{1}{2}\frac{d\ln Z_{3}}{d\ln\mu}=\frac{e^{2}}{48\pi^{2}}
-\frac{e^{2} u^{2}Q_{c}}{64\pi^{2}} \\
\gamma_{m} & = & \frac{1}{m}\frac{d m}{d\ln\mu}=-\frac{2e^{2}-\lambda}{16 \pi^{2}}
+\frac{\lambda u^{2}Q_{c}}{16\pi^{2}},
\end{eqnarray}
corresponding to the fact that if $u^{2}\neq0$, the scalar field in the Lagrange density is no longer
canonically normalized, whereas for  $u^{2}=0$ these functions are equal to the usual ones,
calculated without Lorentz violation. The consequences of relaxing the tracelessness property of
$c^{\mu\nu}$ and $\kappa^{\mu\nu\rho\sigma}$ are potentially more interesting for $\beta(e)$, since
in that case we must add the term $-\frac{e^{2}u^{2} Q_{c}}{96\pi^{2}}$ to the result (\ref{betae}).
For a timelike aether-like vector with $u^{2}=1$, along with $Q_{c}>0$, it appears that a non-trivial
fixed point for $\beta(e)$ may arise out of the LV interactions, at $e_{*}=\frac{Q_{c}}{2}$. This is
suggestive, and it contrasts sharply with the $u^{2}=0$ case we have mostly concentrated on---in which
the Lorentz violation does not contribute to the RG running of the electric charge at one-loop order.
In any case, we anticipate that that our results may be useful for more detailed futute studies of
RG behavior in the SME gauge and scalar sectors.

\section{Conclusion}

\label{sec-concl}

This work essentially completes the
one-loop renormalization of the SME's scalar QED sector, to first order in the
CPT-even LV terms in the scalar ($c^{\mu\nu}$) and gauge ($\kappa^{\mu\nu\rho\sigma}$) sectors.
These results broaden our understanding of the perturbative structure of a frequently-neglected
corner of the SME. Specifically, we have calculated the three- and four-point gauge-scalar vertex corrections
in this model. For both functions, we have confirmed that the results, when combined with previously
calculated two-point functions, yield correctly proportionate contributions to
the gauge-invariant structure $k^{\mu\nu}(D_{\mu}\phi)^{*}(D_{\nu}\phi)$ in the effective Lagrange density.

The radiatively-generated quantity $k^{\mu\nu}$ receives contributions 
proportional to both $c^{\mu\nu}$ and $\kappa^{\mu\rho\nu}{}_{\rho}$. As expected in a renormalizable
theory with dimensionless couplings, both sets of radiative corrections are formally logarithmically
divergent, so they must be regulated and renormalized. (While in theories with fermionic loops, some of
the similar contributions may vanish due to their Dirac matrix structures forcing certain traces to be
zero, such a cancellation mechanism clearly cannot work in purely bosonic theories.)
Furthermore, using our calculations of
the renormalization constants of the model, we were also able to calculate RG functions for the LV operators.  

It is interesting to note that corrections to the scalar four-point function
$\langle(\phi^*\phi)^{2}\rangle$ at first order in $c^{\mu\nu}$ and $\kappa^{\mu\nu\rho\sigma}$ 
are actually zero, because the tensors are taken to be traceless.
Moreover, if the background tensors obey a particular relation, namely
$4c^{\mu\nu}+\kappa^{\mu\rho\nu}{}_{\rho}=0$, all the divergent contributions to
$k^{\mu\nu}(D_{\mu}\phi)^{*}(D_{\nu}\phi)$ cancel out. In this case, all the one-loop radiative
corrections to the ``covariantized'' kinetic term in the scalar field Lagrange density are finite.
This resembles the situation discussed in Ref.~\cite{MarHD}, in which the 
divergent contributions to an effective CFJ term also turned out to vanish if
the SME parameters involved satisfied a special relation.

The natural continuations of this study could consist of: first,
a more detailed evaluation of finite radiative corrections, which can contribute to cross sections and
similar quantities in LV scalar QED; second, development of higher-order calculations, including both
higher-loop Feynman diagrams and calculations at second and higher orders in $c^{\mu\nu}$ and 
$\kappa^{\mu\nu\rho\sigma}$; third, inclusion of the possibility of spontaneous gauge symmetry breaking;
and fourth, extension of these results to LV theories of non-Abelian
gauge fields coupled to scalar matter. We intend to undertake further studies in these directions
in subsequent papers. In particular, the non-Abelian version of the theory represents one of the last
remaining components of the minimal SME whose one-loop renormalization has not yet been completed, although
some of the necessary calculations are fairly straightforward generalization of ones that have already
been done. For example, to generalize the three-point scalar-vector diagrams shown in figure~\ref{Fig1},
it is only necessary to
keep track of internal non-Abelian group generators at the vertices and to include diagrams in which
the external vector line is attached to the internal gauge propagator with a three-gauge-boson vertex.
Ultimately,
all of these quantum correction calculations will enhance our understanding of possible Lorentz violation in
scalar field dynamics, including
elucidating possible experimental signatures of Higgs-sector Lorentz violation.

\section*{Acknowledgments}

The authors are grateful to J. R. Nascimento for important discussions.
The work of A. Yu.\ P. has been partially supported by the CNPq project No. 301562/2019-9.

\end{document}